\documentclass[12pt]{article}
\usepackage{amssymb,amsmath}
\usepackage{epsfig}
\def\shiftleft#1{#1\llap{#1\hskip 0.04em}}
\def\shiftdown#1{#1\llap{\lower.04ex\hbox{#1}}}
\def\thick#1{\shiftdown{\shiftleft{#1}}}
\def\b#1{\thick{\hbox{$#1$}}}

\begin{document}
\title{Non-reactive scattering of excited exotic hydrogen atoms)\thanks{The work was
supported by Russian Foundation for Basic Research (grant No. 03-02-16616).}}

\author{G.Ya. Korenman, V.P. Popov and V.N. Pomerantsev \and
{\em Institute of Nuclear Physics, Moscow State University}}
\date{}
\maketitle

\begin{abstract} The Coulomb deexcitation of light exotic
atoms in collisions with hydrogen atoms has been studied in
the framework of the fully quantum-mechanical
close-coupling method for the first time. The calculations
of the $l$-averaged cross sections have been performed for
$(\mu p)_n$ and  $(\mu d)_n$ atoms in the states with the
principal quantum number $n=3 \div 8$ and relative energies
region $E=0.01 \div 100$~eV. The obtained results reveal
the new $n$ and $E$ dependences of the Coulomb deexcitation
cross sections. The large fraction (up to $\sim$ 36$\%$) of
the transition with $\Delta n > 1$  is also predicted.
\end{abstract}
\section{Introduction}
In this paper we present {\em ab initio} quantum-mechanical treatment
of non-reactive
scattering processes of the excited exotic hydrogen-like atoms:
\begin{itemize}
\item[-] elastic scattering
\begin{equation}
(ax)_n + (be^-)_{\nu} \to (ax)_n + (be^-)_{\nu};
\end{equation}
\item[-] Stark mixing
\begin{equation}
(ax)_{nl} + (be^-)_{\nu} \to (ax)_{nl'} + (be^-)_{\nu};
\end{equation}
\item[-] Coulomb deexcitation
\begin{equation}
(ax)_{nl} + (be^-)_{\nu} \to (ax)_{n'l'} + (b^-e)_{\nu}.
\end{equation}
\end{itemize}
Here $(a,b) = (p,d,t)$ are hydrogen isotopes and $x=\mu^-,
\pi^-,K^-, \tilde{p}$; $(n,l)$ are the principal and
orbital quantum numbers of exotic atom and $\nu$ are the
hydrogen atom quantum numbers. The processes (1) - (2)
decelerate while Coulomb deexcitation (3) accelerates the
exotic atoms, influencing their quantum number and energy
distributions during the cascade. The last process has
attracted  particular attention especially after the "hot"
$\pi p$ atoms with the kinetic energy up to 200 eV were
found experimentally [1,2].

 Starting from the classical paper by Leon and Bethe~[3],
 Stark transitions has been treated in the semiclassical
 straight-line-trajectory approximation (see [4] and
 references therein). The fully quantum-mechanical treatment
 of the processes (1) - (2) based on the adiabatic description
 was given in [5-8]. Recently [9,10], the elastic scattering
 and Stark transitions have also been studied in a close-coupling
 approach with the electron screening effect taken into account by the
model. Thus, at the present time we have more or less
realistic description of the processes (1)-(2) in a wide
range of $n$ and $E$.

Concerning the acceleration process (3), the situation here
is much less defined, especially for low $n$. The first
work (to the best of our knowledge) on the CD process was
performed by Bracci and Fiorentini [11] using the
semiclassical approach with some additional approximations
concerning the assumption of the large values of the
quantum numbers under consideration ($n\gg 3$) and some
others. In the following numerous papers [12, 13] (and
references therein) the CD process is considered within the
asymptotic approaches using the adiabatic hidden crossing
theory [14] for the pure Coulomb three-body systems or for
the screened Coulomb centers chosen in the static form (the
dipole screening). Recently, the calculations of the
process (3) were also performed in the classical-trajectory
Monte-Carlo (CTMC) [15] approach. While the Coulomb
deexcitation cross sections obtained in CTMC approach are
in fair agreement with the semiclassical ones of Bracci and
Fiorentini [11], the much more elaborated advanced
adiabatic approach (AAA) [12,13] gives much smaller Coulomb
deexcitation cross sections which can not explain the
experimental data. The reasons of such a strong discrepancy
are not clear. One can only assume that the semiclassical
model [11]  as well as the CTMC approach are not valid for
low-lying states. Thus, until now there is no satisfactory
description of this process in the most interesting region
($n=3 \div7 $).

The main aim of this paper is to obtain the cross sections
of Coulomb deexcitation for $3\le n \le 8$ in realistic
quantum-mechanical approach which is free from the
additional approximations used in previous studies, in
particular, the two-state approximation.  For this reason
we use a unified treatment of elastic scattering, Stark
transitions and Coulomb deexcitation within the
close-coupling method. This approach was used previously
for the study of the elastic scattering and Stark
transitions of the excited muonic hydrogen in the
collisions with the hydrogen ones~[16] and a good agreement
as compared with the results of the quantum-mechanical
adiabatic description~[5-8] was obtained.
 In the following section we briefly describe the
 close-coupling formalism and the main expressions.
 The results of the close-coupling  calculations concerning
 the total cross sections of the process (3) are presented
 and discussed in Sec. III. Finally, summary and concluding
 remarks are given in Sec. IV.

\section{Theory}
 The non-relativistic Hamiltonian for the four-body system
$(a \mu^{-} + b e^{-})$, after separating the center of
mass motion, can be written in Jacobi coordinates
$(\mathbf{R}, \boldsymbol{\rho}, \mathbf{r})$ as
\begin{equation}
H = -\frac{1}{2m}\Delta _{\mathbf{R}} +
h_{\mu}(\boldsymbol{\rho}) + h_e(\mathbf{r})
+V(\mathbf{r},\boldsymbol{\rho},\mathbf{R})
 \label{Ham},
\end{equation}
where $m$ is the reduced mass of the system, $V$ is the
electrostatic interaction between the subsystems, $h_\mu $
and $h_e $ are the Hamiltonian of the free exotic and
hydrogen atoms
\begin{align}
h_\mu \Phi_{nlm}(\boldsymbol{\rho}) &= \varepsilon_n
\Phi_{nlm}(\boldsymbol{\rho}),
\\
h_e\varphi_{1s}(\mathbf{r}) &=
\epsilon_{1s}\varphi_{1s}(\mathbf{r}),
\end{align}
where $\Phi_{nlm}(\boldsymbol{\rho})$ and
$\varphi_{1s}(\mathbf{r})$ are the wave functions of the
exotic atom and hydrogen atom bound states, $\varepsilon_n$
and $\epsilon_{1s}$ are the corresponding eigenvalues. The
interaction potential of the subsystems is defined by
\begin{equation}
V(\mathbf{r},\boldsymbol{\rho},\mathbf{R})=V_{ab}+V_{\mu
b}+V_{ae}+V_{\mu e} \label{}
\end{equation}
with the two-body Coulomb interactions
\begin{eqnarray}
 V_{ab}=\frac{1}{r_{ab}}=
 |\mathbf{R}+\nu \boldsymbol{\rho}-\nu_e \mathbf{r}|^{-1},
 V_{\mu b}=
 -\frac{1}{r_{\mu b}}=
 -|\mathbf{R}-\xi\boldsymbol{\rho}-\nu_e \mathbf{r}|^{-1},
 \\
 V_{\mu e}=
 \frac{1}{r_{\mu e}}=
 |\mathbf{R}-\xi \boldsymbol{\rho}+\xi_e \mathbf{r}|^{-1},
V_{ae}=-\frac{1}{r_{ae}}=-|\mathbf{R}+\nu
\boldsymbol{\rho}+\xi_e \mathbf{r}|^{-1}.
\end{eqnarray}
Here we use the set of Jacobi coordinates $(\mathbf{R},
\boldsymbol{\rho},  \mathbf{r})$:
\[\mathbf{R} =
\mathbf{R}_H - \mathbf{R}_{\mu a},\quad \boldsymbol{\rho} =
\mathbf{r}_\mu - \mathbf{r}_a,\quad \mathbf{r}=
\mathbf{r}_e -\mathbf{r}_b,\] where $\mathbf{r}_a,
\mathbf{r}_b, \mathbf{r}_\mu, \mathbf{r}_e$ are the
radius-vectors of the nuclei, muon and electron in the
lab-system , $\mathbf{R}_H, \mathbf{R}_{\mu a}$ are the
center of mass radius-vectors of the hydrogen and exotic
atoms, respectively. The coefficients $\nu$, $\xi$, $\nu_e$ and
$\xi_e$ in the two-body interactions (8)-(9) depend on the
masses of the particles,
\begin{eqnarray}
\nu = m_\mu /(m_\mu +m_a),\; \xi = m_a /(m_\mu +m_a),
\\ \nu_e = m_e /(m_e+m_b), \; \xi_e = m_b /(m_e +m_b),
\end{eqnarray}
and satisfy
\begin{equation}
\nu + \xi = \nu_e + \xi_e = 1. \label{}
\end{equation}
($m_a, m_b, m_{\mu}$ and $m_e$ are the masses of the
hydrogen isotopes, muon and electron, respectively).
 Atomic units (a.u)
 $\hbar=e=m_e m_b/(m_e+m_b)=1$ will be used throughout the
 paper unless otherwise stated.\\

The total wave function $\Psi (\boldsymbol{\rho},
\mathbf{r}, \mathbf{R})$ of the system satisfies the time
independent Schr\"{o}dinger equation with the Hamiltonian
(4):
\begin{equation}
( -\frac{1}{2m}\Delta _{\mathbf{R}} + h_\mu + h_e +V)\Psi
(\boldsymbol{\rho},  \mathbf{r}, \mathbf{R}) = E\Psi
(\boldsymbol{\rho},  \mathbf{r}, \mathbf{R}) , \label{}
\end{equation}
where $E$ is the total energy of the system.

  In this paper, as well as in the previous studies [11, 12, 15],
   we assume that the state of the target
 electron is fixed during the collision. The electron
 excitations can be taken into account in a straightforward
 manner.  Owing to the rotation and inversion symmetries of the total
 system, the vector solutions
 of (13) is introduced the total angular momentum representation.
In a space-fixed coordinate frame we built the basis states
from the eigenvectors of the operators  $ h_e, h_\mu,
\mathbf{l}^2, \mathbf{L}^2, \mathbf{J}^2, J_z$ and the
total parity $\pi$ with eigenvalues $\varepsilon_{1s},
\varepsilon_n, l(l+1), L(L+1), J(J+1), M$ and $(-1)^{l+L}$,
respectively:
\begin{equation}
|\Gamma \rangle \equiv \phi_{1s}(\mathbf{r})|\gamma\rangle
\label{}
\end{equation}
where
\begin{equation}
|\gamma \rangle \equiv|n l, L:JM\rangle\equiv i^L \sum_{m
\lambda}\langle l m
L\lambda|JM\rangle\Phi_{nlm}({\b\rho}) Y_{L
\lambda}(\mathbf{\hat R}),  \label{}
\end{equation}
\begin{equation}
|\Gamma \rangle \equiv \frac{1}{\sqrt{4\pi}} R_{1s}(r)
R_{nl}(\rho) {\cal Y}_{lL}^{JM}(\hat {\b\rho}, \hat{\bf
R}), \label{}
\end{equation}
\begin{equation}
{\cal Y}_{l L}^{JM} (\hat {\b\rho}, \hat{\bf R})\equiv
 i^{l+L}\sum_{m \lambda}\langle l m L \lambda |JM
\rangle Y_{lm}(\hat{\b\rho}) Y_{L \lambda}(\mathbf{\hat R})
 \label{}
\end{equation}
Here the orbital angular momentum $\bf l$ of $(a\mu)_{nl}$
is coupled with the orbital momentum $\bf L$ of the
relative motion to give the total angular momentum, $\bf
{J=l + L}$.  Then, for the fixed values of $J, M, \pi =
(-1)^{l+L}$ the exact solution of the Schr\"{o}dinger
equation for the colliding system,
\begin{equation}
(E - H) \Psi_{E}^{JM\pi}({\bf r}, {\b\rho}, {\bf
R}) = 0,   \label{}
\end{equation}
is expanded as follows
\begin{equation}
\Psi_{E}^{J M \pi}(\mathbf{r}, {\b\rho},
\mathbf{R}) = \varphi_{1s}({\bf r})\frac{1}{R} \sum_{nl
L}G_{nlL}^{J \pi}(R)|nl,L:JM\rangle, \label{}
\end{equation}
where the $G_{nlL}^{J \pi}(R)$ are the radial channel
functions and the sum is restricted to $(l, L)$ values to
satisfy the total parity conservation.  This expansion
leads to the coupled radial scattering equations
\begin{equation}
\left(\frac{d^2}{dR^2} + k^2_{n} -
\frac{L(L+1)}{R^2}\right)G^{J \pi}_{nlL}(R) = 2m
\sum_{n'l'L'}W^{J}_{nlL,n'l'L'}(R)\,G^{J \pi }_{n'l'L'}(R),
\label{cce}
\end{equation}
where
$k^{2}_{n}=2m(E-\varepsilon_{n}-\epsilon_{1s})=2m(E_{cm}+\Delta\varepsilon_{n\,n'})$
specify the channel wave numbers; $E_{cm}$ is
the relative motion energy in the entrance channel,
$\Delta\varepsilon_{n\,n'}=0.5\mu(n^2-(n')^2)/(n n')^2$ is
the difference of the exotic atom bound energies in the
initial and the final states, $\mu$ is the reduced mass of
the exotic atom. Finally,  $W^{J}_{nlL,n'l'L'}$ are the
matrix elements that couple asymptotic channels $(n l L;
J)$ and $(n' l' L'; J)$:
\begin{equation}
W^{J}_{nlL,n'l'L'}\equiv \langle
1s,nl,L:J|\hat{V}|1s,n'l',L':J\rangle.
\end{equation}
The radial functions $G_{E, n'l'L'}^{J \pi}(R)$ must be
regular everywhere, and, at $R\rightarrow 0$
\begin{equation}
G_{E, n'l'L'}^{J \pi}(0)=0 (\sim R^{L+1}) \label{}
\end{equation}
and at asymptotic distances ($R\rightarrow \infty$) satisfy
the usual boundary conditions
\begin{equation}
G_{E, n'l'L'}^{J \pi}(R)\Rightarrow \frac{1}{\sqrt
{k_f}}\{\delta_{if} \delta_{nn'} \delta_{ll'}
\delta_{LL'}e^{-i(k_{i}R-L\pi/2)}- S^J(nl, L\rightarrow
n'l', L')e^{i(k_{f}R-L'\pi/2)}\}, \label{}
\end{equation}
where $k_i$, $k_f$ are the wave numbers of initial and
final channels and $S^J(nl, L\rightarrow n'l', L')$ is the
scattering matrix in the total angular momentum
representation. The indexes of the entrance channel and
target electron state are omitted for brevity. The
scattering amplitude for $nlm \rightarrow n'l'm'$ is
defined by
\begin{align}
f(nlmL\rightarrow n'l'm'L'|\mathbf{k}_i, k_f, \mathbf {\hat R}\rangle&=
\frac{2\pi i}{\sqrt{k_{i}k_{f}}}\sum_{JMLL'\lambda
\lambda'}i^{L'-L}\langle lmL\lambda |J M\rangle \langle l'm'L'\lambda'|J M\rangle
\times \nonumber \\
&\times Y_{L\lambda}^{*}(\mathbf{\hat k_i})
Y_{L'\lambda'}(\mathbf{\hat R})T^J(nlL\rightarrow n'l'L'),
\label{}
\end{align}
where the transition matrix $T^J$ used here is given by
\begin{equation}
T^J(nl, L\rightarrow n'l',
L')=\delta_{nn'}\delta_{ll'}\delta_{LL'}\delta_{mm'}\delta_{\lambda\lambda'}
- S^J(nl, L\rightarrow n'l', L'). \label{}
\end{equation}

The matrix elements (21) of the interaction potential (7-9)
are given by
\begin{align}
W^{J}_{nlL,n'l'L'}(R)&=\frac{1}{4\pi}\int{\rm d} {\bf
r}\,{\rm d}{\b \rho} \,{\rm d} \hat{\bf R} R^2_{1s}
(r)R_{nl} (\rho)R_{n'l'}(\rho) \nonumber\\ &\times ({\cal
Y}^{JM}_{lL})^{^*} (\hat{\b \rho},\hat{\bf R})\, V({\bf
r},{\b \rho},{\bf R})\, {\cal Y}^{JM}_{l'L'} (\hat{\b
\rho},\hat{\bf R}),
\end{align}
 where the radial hydrogen-like wave functions
 are given explicitly  by
\begin{equation}
R_{nl}(\rho)=N_{nl}\left(\frac{2\rho}{n
a}\right)^l\exp(-\rho/na)
 \sum_{q=0}^{n-l-1}S_{q}(n,
 l)\left(\frac{2\rho}{n a}\right)^q \label{}
\end{equation}
($a$ is the Bohr' radius of the exotic atom in a.u.) with
\begin{equation}
N_{nl}=\left(\frac{2}{n a}\right)^{3/2}
\left[\frac{(n+l)!(n-l-1)!}{2n}\right]^{1/2}, \label{}
\end{equation}
\begin{equation}
S_{q}(n, l) = (-)^q \frac{1}{q!(n-l-1-q)!(2l+1+q)!}.
\label{}
\end{equation}
 Averaging $V({\bf r},{\b \rho},{\bf R})$ in (26) over $1s$-state of hydrogen atom
 we obtain
\begin{align}
V({\bf R},{\b \rho})&=\frac{1}{4\pi}\int_{0}^{\infty}{\rm d}
{\bf r} R^2_{1s} (r)V({\bf r},{\b \rho},{\bf R})=
\nonumber\\
&= \frac{1}{\xi_e}\{U_{\nu,\xi_e}({\bf
 R},{\b \rho})-U_{-\xi,\xi_e}({\bf R},{\b \rho})\} -
\frac{1}{\nu_e}\{U_{\nu,\nu_e}({\bf R},{\b
\rho})-U_{-\xi,\xi_e}({\bf R},{\b \rho})\},
\end{align}
where
\begin{multline}
U_{\alpha,\beta}({\bf R}, {\b \rho})=(1+\frac{\beta}{|{\bf R}+
\alpha\boldsymbol{\rho}|}) {\rm e}^{-\frac{2|{\bf R}+\alpha\boldsymbol{\rho}|}{\beta}}\equiv
 \lim_{x\to 1}\left(1-\frac{1}{2}\frac{\partial}{\partial x}\right)
\beta\frac{{\rm e}^{-\frac{2x|{\bf R}+\alpha\boldsymbol{\rho}|}{\beta}}} {|{\bf
R}+\alpha\boldsymbol{\rho}|}.
\end{multline}
Using the additional theorem for the spherical Bessel
functions
\begin{multline}
\frac{{\rm e}^{-\lambda |{\bf R}_1+{\bf r}_1|}}{|{\bf R}_1+{\bf r}_1|}=
\frac{4\pi}{\sqrt{R_1r_1}}\sum_{t\tau}(-1)^t Y^*_{t\tau}(\hat{\bf R}_1)
Y_{t\tau}(\hat{\bf r}_1)\times
\\ \times \left\{ K_{t+1/2}(\lambda R_1)\,I_{t+1/2}(\lambda
r_1)\left |_{r_1 <R_1} + I_{t+1/2}(\lambda R_1)\,K_{t+1/2}(\lambda r_1)\right
|_{r_1 >R_1} \right \}
\end{multline}
($I_p(x)$ and $K_p(x)$ are the modified  spherical Bessel functions of the first
and third kind) and substituting eqs.(30-32) into (26), we integrate over the angular
variables. Furthermore, using
 the angular momentum algebra and integrating over $\rho$, we obtain:
\begin{align}
W^{J}_{nlL,n'l'L'}(R)&=(-1)^{J+l+l'}i^{l'+L'-l-L}\sqrt{\hat{l}\hat{l'}\hat{L}\hat{L'}}
\sum_{t=0}^{t_m}(l0l'0|t0)(L0L'0|t0)
\left\{\begin{array}{lll}
l&l'&t\\L'&L&J\end{array}\right\}\times \nonumber \\
&\times \left\{\frac{1}{\xi_e}\left[ (-1)^t W_t(R,\nu
,\xi_e;nl,n'l') - W_t (R,\xi,\xi_e;nl,n'l')\right] \right
.- \nonumber \\ & - \left .  \frac{1}{\nu_e}\left[ (-1)^t
W_t(R,\nu,\nu_e;nl,n'l') - W_t(R,\xi,\nu_e;nl,n'l')\right]
\right\}
\end{align}
($t_m$ is the maximum value of the allowed multipoles).
Here the next definitions are used:
\begin{align}
W_t(R,\alpha,\beta;n l,n'l')&={\cal N}_{nl,n'l'}
\sum_{m_1=0}^{n-l-1}S_{m_1}(n,l)\left(\frac{2n'}{n+n'}\right)^{m_1}
\sum_{m_2=0}^{n'-l'-1} S_{m_2}(n', l')\left(\frac{2n}{n+n'}\right)^{m_2} \times
\nonumber \\
\times &\left\{H_t(z)J_1^{t,s}(z,\lambda(n,n',\alpha,\beta))-
h_t(z)J_2^{t,s}(z,\lambda(n,n',\alpha,\beta)) + \right . \nonumber \\
&+F_t(z)J_3^{t,s}(z,\lambda(n,n',\alpha,\beta)) \left .
+f_t(z)J_4^{t,s}(z,\lambda(n,n',\alpha,\beta)) \right\},
\end{align}
where $z=2R/\beta $, $ s=l+l'+m_1+m_2 $, $ \hat{L}\equiv 2L+1$;
\begin{equation}
{\cal N}_{nl,n'l'} =
\frac{1}{n+n'}\!\left(\frac{2n'}{n+n'}\right)^{l+1}\!\left(\frac{2n}{n+n'}\right)^{l'+1}
\!\sqrt{(n+l)!(n-l-1)!(n'+l')!(n'-l'-1)}; \label{}
\end{equation}
\begin{equation}
\lambda(n,n',\alpha,\beta)=\frac{2nn'}{n+n'}\frac{a\alpha}{\beta};
\label{}
\end{equation}
\begin{equation}
H_t(x)=(1-2t)h_t(x)+xh_{t+1}(x); \label{}
\end{equation}
\begin{equation} F_t(x)=(1-2t)f_t(x)-xf_{t+1}(x). \label{}
\end{equation}
The functions $h_t(y)$ and $f_t(y)$ are given by
\begin{equation}
h_t(y)\equiv\sqrt{\frac{2}{\pi y}}K_{t+1/2}(y) \label{}
\end{equation}
and
\begin{equation}
f_t(y)\equiv\sqrt{\frac{\pi}{2 y}}I_{t+1/2}(y).
\label{}
\end{equation}

The radial integrals $J_i^{t,s}(x,\lambda)$ are defined as follows:
\begin{equation}
J_1^{t,s}(x,\lambda)=\int_{0}^{x/\lambda}
y^{s+2}e^{-y}f_t(\lambda y)\,{\rm d}y,
\end{equation}
\begin{equation}
J_2^{t,s}(x,\lambda)=\lambda J_1^{t+1,s+1}(x,\lambda),
\end{equation}
\begin{equation}
J_3^{t,s}(x,\lambda)=\int_{x/\lambda}^{\infty}
y^{s+2}e^{-y}h_t(\lambda y)\,{\rm d}y,
\end{equation}
\begin{equation}
J_4^{t,s}(x,\lambda)=\lambda J_3^{t+1,s+1}(x,\lambda)
\end{equation}
and calculated analytically using  the power series
 for the modified Bessel functions.

Finally, we give the explicit expressions for the cross
sections to be discussed in this paper. The partial cross
sections of the processes (1) - (3) for the transitions $n
l \rightarrow n'l'$, averaging over an initial distribution
of the degenerated substates and summed over the degenerate
final substates, are given by
\begin{equation}
\sigma^J(n l \rightarrow n'l'; E) =
\frac{\pi}{k_{i}^2}\frac{2J+1}{2l+1} \sum_{L
L'}|T^J(nlL\rightarrow n'l'L')|^2. \label{}
\end{equation}

The $l$-averaged partial cross sections for the transitions
$n \rightarrow  n'$ are then computed by summing over $l$
and $l'$ with the statistic factor $(2l+1)/n^2$:
\begin{equation}
\sigma_{n\to n'}^J(E) = \frac{\pi}{k_{i}^2}\frac{2J+1}{n^2}
\sum_{l,\,l\,' \, L L'}|T^J(nlL\rightarrow
n'l'L')|^2. \label{}
\end{equation}
The total cross sections for the transition $n \rightarrow
n'$ is obtained by summing the corresponding partial cross
section over the total angular momentum $J$:
\begin{equation}
\sigma_{nn'}(E) = \sum_{J}\sigma^J_{nn'}(E).
\end{equation}
For the discussion of the obtained results and the comparison
of them with the other approaches we will also need the
total cross section of Coulomb deexcitation including all
transitions $n\to n'$ with $n'<n$:
\begin{equation}
\sigma^{\rm CD}_n(E) = \sum_{n'<n}\sigma_{nn'}(E).
\end{equation}

\section{Results}
The close-coupling method described in the previous Section
has been used to obtain the cross sections for the
collisions of the $\mu^-p$ and $\mu^-d$ atoms in excited
states with hydrogen atoms. The present paper had at least
two goals: first, to apply the fully quantum-mechanical
approach for the study of the processes (1) - (3) and,
second, to clear the validity of the energy and principle
quantum number dependence of the Coulomb deexcitation cross
sections used in literature and based on the semiclassical
model~[11]. So, we present here only a small part of our
results. In particular, to illustrate some main features of
the calculated cross sections, we discuss only the
$l$-averaged CD cross sections. The detailed results of the
calculations will be published elsewhere.

The coupled differential equations (20) are solved
numerically using the Numerov method and with the real
boundary conditions involving the $K$-matrix instead of
$S$-matrix. The corresponding $T$-matrix can be calculated
from equation
\[ T=2K(I- i K)^{-1}=2K^2(1+K^2)^{-1}-2iK(1+K^2)^{-1}, \]
where $I$ is the unit matrix. In the calculations we have
taken into account all possible multipoles in the
interaction potentials and all open channels with $n'\le
n$.

The close-coupling calculations have been carried out for
the relative collision energies $E_{\rm cm}$ from 0.01 up
to 100 eV and for the excited states with $n=3\div 8$. At
all energies we take into account so many values of the
total angular momentum $J$ and the relative angular
momentum $L$ that all the inelastic cross sections were
calculated with the accuracy not less than 0.1\%.

\begin{table}[h]
\caption{Dependence of the elastic and deexcitation cross
sections $\sigma_{nn'}$ for $(\mu d)_n + {\rm H}$
collisions at energy $E_{cm}=0.1$~eV on the number $t_m$ of
terms included in the multipole expansion for the
matrix elements (33).}
\begin{center}
\begin{tabular}{|c|c|c|c|c|c|c|} \hline
$t_m$ & $\sigma_{66}$ & $\sigma_{65}$& $\sigma_{64}$& $\sigma_{63}$&
$\sigma_{62}$& $\sigma_{61}$ \\ \hline

  1   &  54.70  &  2.36  &  0.20 &   0.012 &   0.0003 &   0.0000002 \\
  2   &  50.24  &  5.55  &  0.37 &   0.055 &   0.0052 &   0.000008  \\
  3   &  51.58  &  4.82  &  0.38 &   0.038 &   0.0176 &   0.000038  \\
  4   &  52.65  &  2.71  &  0.38 &   0.031 &   0.0177 &   0.000060  \\
  5   &  52.31  &  2.72  &  0.39 &   0.026 &   0.0277 &   0.000079  \\
  6   &  54.19  &  1.42  &  0.36 &   0.031 &   0.0295 &   0.000093  \\
  7   &  54.77  &  1.04  &  0.32 &   0.032 &   0.0279 &   0.000100  \\
  8   &  54.46  &  1.18  &  0.30 &   0.030 &   0.0270 &   0.000102  \\
  9   &  54.36  &  1.17  &  0.29 &   0.028 &   0.0269 &   0.000103  \\
 10   &  54.31  &  1.20  &  0.30 &   0.029 &   0.0271 &   0.000103  \\ \hline
 \end{tabular}
 \end{center}
\end{table}

The results of the calculations are presented in Tables 1-3
and Figs. 1-6. The effects of the higher multipoles in the
expansion of the interaction potentials are large (see
Table 1). In contrast to the elastic scattering and Stark
transitions in which the ``dipole'' approximation ($t_m=1$)
gives practically the exact results, there is no case for
the Coulomb deexcitation. Ignoring the higher multipoles
would lead to the significant distortions of the cross
sections. As it is seen from Table~1 the results are
sharply varied in the top part of the table ($t_m=1\div
7$) and are practically unchanged in the bottom one
($t_m=8\div 10$). Thus, to provide the proper treatment of
the Coulomb deexcitation all the multipoles in the
expansion of the interaction potential must be taken into
account.

In Fig.~1 we present the $J$ dependence of the partial-wave
cross sections $\sigma^J_{nn'}$ for $n=7$ at three fixed
energies 0.1, 2, and 50 eV. It is seen that a substantial
part of the Coulomb cross sections comes from the partial
waves with rather a low $J$ (in comparison with the elastic
and Stark mixing processes). It is worthwhile to note that
for $n=3$ the range of $J$ values is much less, e.g. at
relative energies 0.01, 2 and 100 eV we obtain $J^{CD}_{\rm
max}=7,11,15$ respectively, whereas for the elastic
scattering we obtain $J_{\rm max}=7,60,120$ for the same
energies.

     \begin{figure}[h]
     \centerline{\includegraphics[width=0.7\textwidth,keepaspectratio]{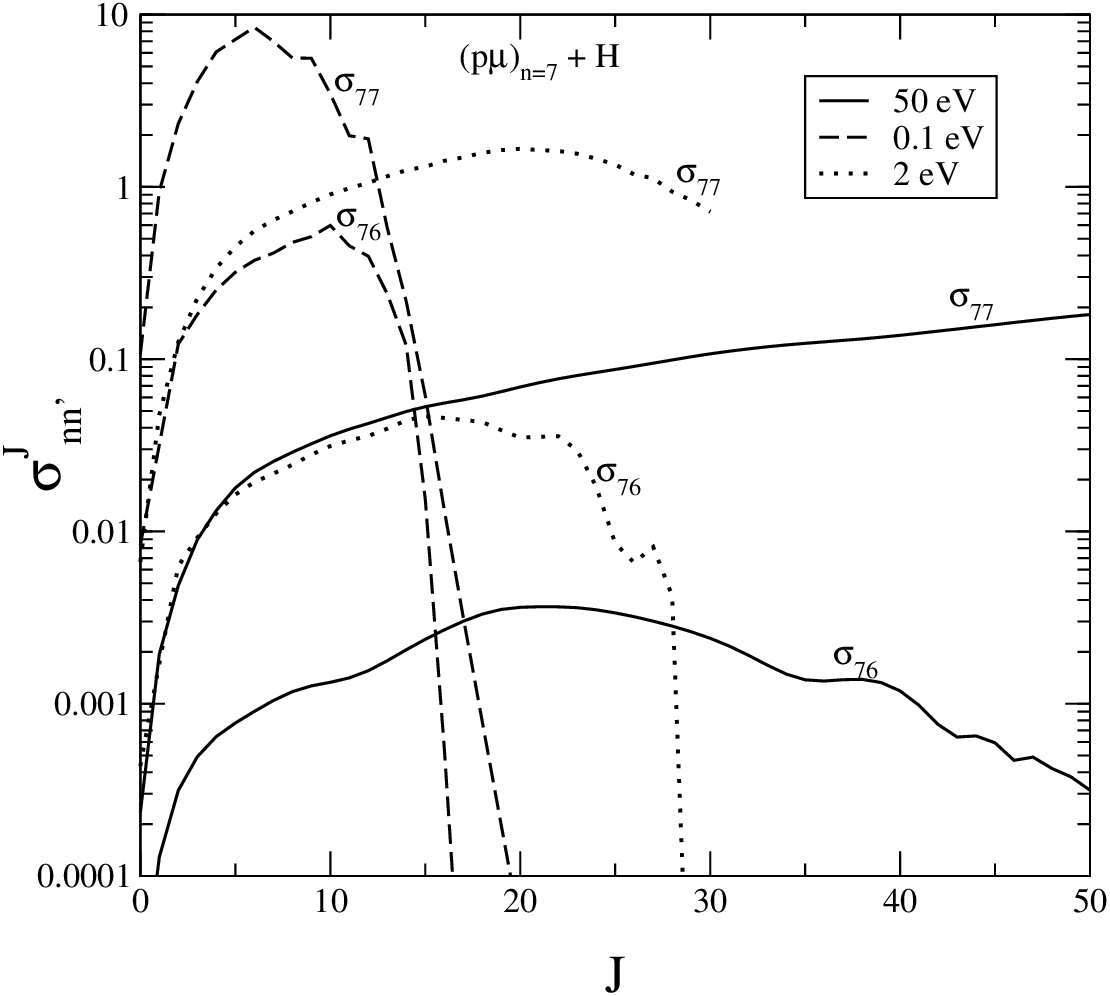}}
     \caption{The partial cross sections of the Coulomb deexcitation
     $\sigma^J_{7,6}$ and the elastic scattering $\sigma^J_{7,7}$ in a.u.
     for $(\mu p)_{n=7} + H$ collisions  as function of the total angular momentum
     $J$ at three energies: 0.1~eV (dashed), 2~eV (dotted) and 50~eV (solid).}
     \end{figure}

The convergence of the results upon the number of open
channels (the transitions with $\Delta n > 1$ are included)
is illustrated in Table~2 for $(\mu p)_{n=7} +D$ collisions
at the relative energy $E_{\rm cm}=1$~eV. It is seen that
only the nearest channels are most important. While $n$
increasing the energy gap between the exotic atom states
decreases. As a result the influence of the nearest
channels becomes more significant and leads to the new $n$
and $E$ dependence of the Coulomb deexcitation cross
sections as compared with the semiclassical ones.

\begin{table}[h]
\caption{The dependence of elastic and Coulomb deexcitation
cross sections $\sigma_{7n'}$ for $(\mu p)_n + {\rm D}$
collisions at energy $E_{cm}=1$~eV on the number of open
channels $N_{\rm level}$ included in the calculation (all
terms of the multipole expansion for the potential are
taken into account).}

\begin{center}
\begin{tabular}{|c|c|c|c|c|c|c|c|} \hline
$N_{\rm level}$ & $\sigma_{77}$ & $\sigma_{76}$&
$\sigma_{75}$& $\sigma_{74}$& $\sigma_{73}$& $\sigma_{72}$
& $\sigma_{71}$\\ \hline

 2 &   39.578016  &  1.063412  &            &            &            &             &            \\
 3 &   40.181324  &  0.679067  &   0.096652 &        &        &         &        \\
 4 &   40.181170  &  0.598721  &   0.056935 &   0.009409 &        &         &        \\
 5 &   40.177221  &  0.598219  &   0.055336 &   0.009368 &   0.001601 &         &        \\
 6 &   40.177061  &  0.597370  &   0.055110 &   0.009343 &   0.001635 &   0.001241  &        \\
 7 &   40.177065  &  0.597360  &   0.055110 &   0.009343 &   0.001635 &   0.001241  &   0.000005 \\
\hline
\end{tabular}
\end{center}
\end{table}

\begin{table}
\caption{The cross sections of Coulomb deexcitation for $(\mu p)_n + H$
 collisions calculated in the quantum-mechanical close-coupling method.}

\begin{tabular}{|c|c|c|c|c|c|c|c|c|c|c|c|c|} \hline
    $E_{\rm cm}$, eV &   0.01 &   0.05 &   0.1 &   0.2 &   0.5 &   1.0 &   2.0 &   5.0 &   7.0 &  10.0 &  15.0 &  20.0   \\ \hline
       $\sigma_{3,2}$ &   4.821 &   1.505 &   1.032 &   0.803 &   0.403 &   0.209 &   0.105 &   0.043 &   0.032 &   0.023 &   0.015 &   0.012  \\
$\sigma^{\rm CD}_3$ &   4.848 &   1.511 &   1.035 &   0.805 &   0.403 &   0.210 &   0.105 &   0.043 &   0.032 &   0.023 &   0.015 &   0.012  \\
       $f_3$, \% &     0.6 &     0.4 &     0.3 &     0.2 &     0.2 &     0.2 &     0.2 &     0.2 &     0.2 &     0.2 &     0.2 &     0.2   \\ \hline
       $\sigma_{4,3}$ &   4.126 &   1.730 &   0.936 &   0.704 &   0.515 &   0.333 &   0.189 &   0.083 &   0.060 &   0.043 &   0.030 &   0.024   \\
$\sigma^{\rm CD}_4$ &   5.526 &   2.656 &   1.397 &   0.965 &   0.637 &   0.403 &   0.226 &   0.098 &   0.071 &   0.052 &   0.036 &   0.029  \\
       $f_4$, \% &    25.3 &    34.9 &    33.0 &    27.0 &    19.1 &    17.2 &    16.4 &    16.0 &    16.2 &    16.3 &    16.5 &    16.7  \\ \hline
       $\sigma_{5,4}$ &   4.364 &   1.349 &   0.943 &   0.574 &   0.306 &   0.211 &   0.147 &   0.070 &   0.054 &   0.041 &   0.031 &   0.026  \\
$\sigma^{\rm CD}_5$ &   4.955 &   1.705 &   1.270 &   0.819 &   0.422 &   0.277 &   0.182 &   0.085 &   0.065 &   0.049 &   0.037 &   0.031  \\
       $f_5$, \% &    11.9 &    20.9 &    25.8 &    30.0 &    27.4 &    23.9 &    19.0 &    17.2 &    16.5 &    16.1 &    15.4 &    14.9  \\ \hline
       $\sigma_{6,5} $&   4.063 &   1.804 &   1.619 &   0.856 &   0.442 &   0.250 &   0.136 &   0.058 &   0.043 &   0.032 &   0.024 &   0.021  \\
$\sigma^{\rm CD}_6$ &   6.246 &   2.598 &   2.166 &   1.189 &   0.661 &   0.387 &   0.211 &   0.090 &   0.067 &   0.050 &   0.038 &   0.033  \\
       $f_6$, \%&    35.0 &    30.6 &    25.3 &    28.0 &    33.1 &    35.4 &    35.5 &    36.0 &    36.5 &    36.8 &    36.8 &    36.1  \\ \hline
       $\sigma_{7,6} $&   5.954 &   2.888 &   2.005 &   1.410 &   0.902 &   0.488 &   0.275 &   0.119 &   0.088 &   0.065 &   0.048 &   0.040  \\
$\sigma^{\rm CD}_7$&   7.819 &   3.544 &   2.454 &   1.762 &   1.076 &   0.591 &   0.340 &   0.148 &   0.111 &   0.083 &   0.062 &   0.053  \\
       $f_7$, \%&    23.9 &    18.5 &    18.3 &    20.0 &    16.2 &    17.5 &    19.3 &    19.9 &    20.4 &    21.2 &    22.3 &    23.4  \\ \hline
       $\sigma_{8,7} $&   5.776 &   3.243 &   2.516 &   1.883 &   1.184 &   0.846 &   0.480 &   0.223 &   0.166 &   0.100 &   0.085 &   0.081  \\
$\sigma^{\rm CD}_8$ &   8.201 &   4.221 &   3.151 &   2.320 &   1.426 &   1.023 &   0.587 &   0.272 &   0.203 &   0.130 &   0.111 &   0.105  \\
       $f_8$, \%&    29.6 &    23.2 &    20.1 &    18.8 &    17.0 &    17.3 &    18.1 &    17.9 &    18.0 &    23.3 &    23.0 &    22.5  \\ \hline

\end{tabular}
\end{table}

The energy dependence of the total Coulomb deexcitation
cross sections $\sigma_n^{CD}(E)$ for $n=3\div 8$ is shown
in Fig.2 in comparison with the results of Bracci and
Fiorentini [11] for $n=3,5,7$ and the CTMC calculations of
Jensen and Markushin~[15] for $n=4\div 8$ and $E_{\rm
cm}=0.5$ and 5~eV. The present results are in poor
agreement with the semiclassical results [11] and CTMC
results [15]. There is only a qualitative agreement for
$n=5$ and energies above 2~eV. The energy dependence of the 
Coulomb deexcitation cross sections in our approach is 
approximately given by $1/\sqrt{E_{\rm cm}}$ at $E_{\rm cm}
\lesssim 0.5$~eV (see Fig.3), in contrast to the
$1/E_{\rm cm}$ behavior found for low energies in [11] and
also in the advanced adiabatic approach (AAA) [13].

     \begin{figure}[h!]
     \centerline{
     \includegraphics[width=0.7\textwidth,keepaspectratio]{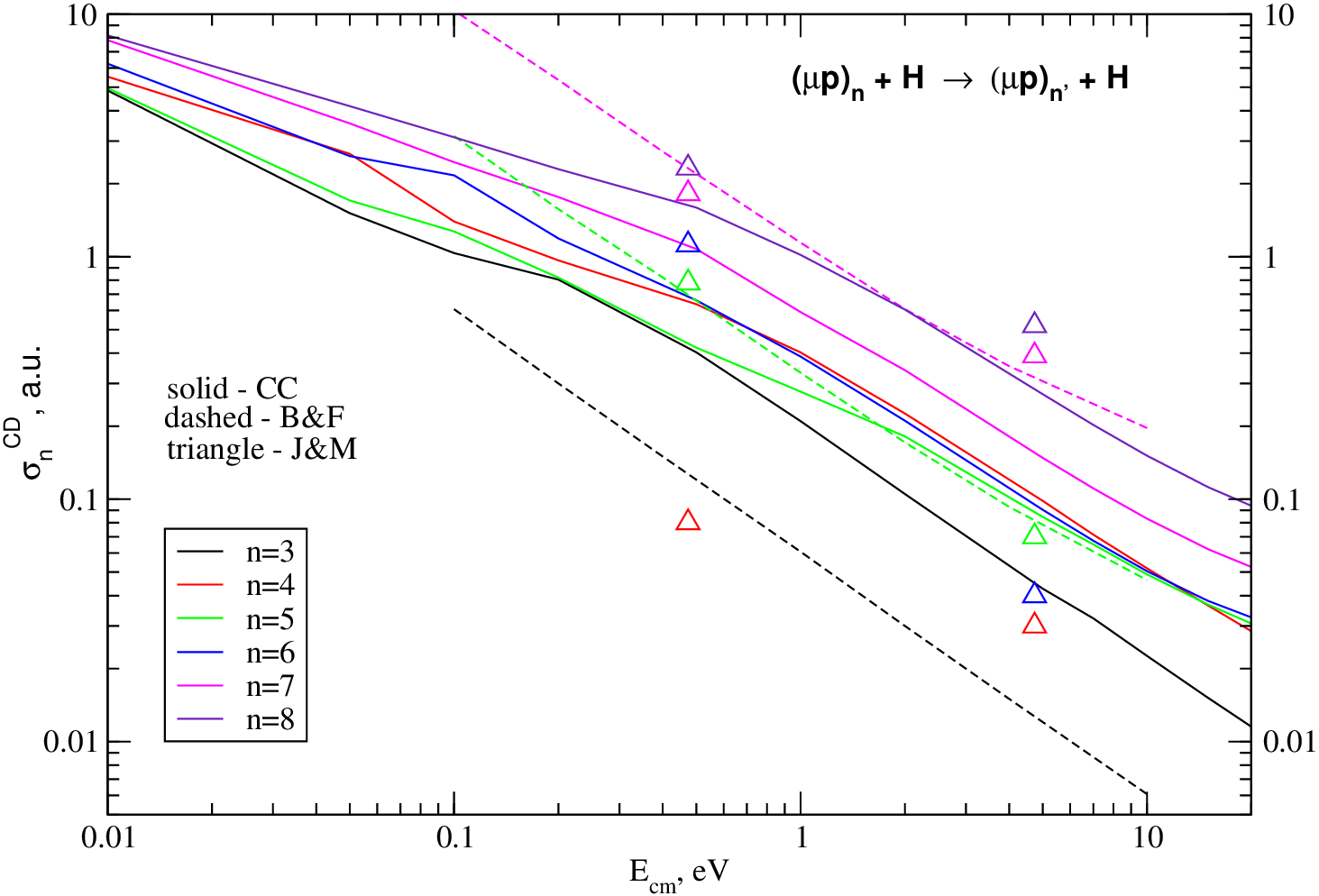}}
     \caption{The cross sections of Coulomb deexcitation
     $\sigma^{\rm CD}_n$ in a.u. for $(\mu p)_n + H$
 collisions calculated in the quantum-mechanical close-coupling method (solid
 lines) in comparison with the results of Bracci and Fiorentini~[11] (dashed) and
 classic Monte-Carlo results~[15] (triangles).}
     \end{figure}

   \begin{figure}[h!]
   \vspace*{0.5 cm}
     \centerline{
     \includegraphics[width=0.6\textwidth,keepaspectratio]{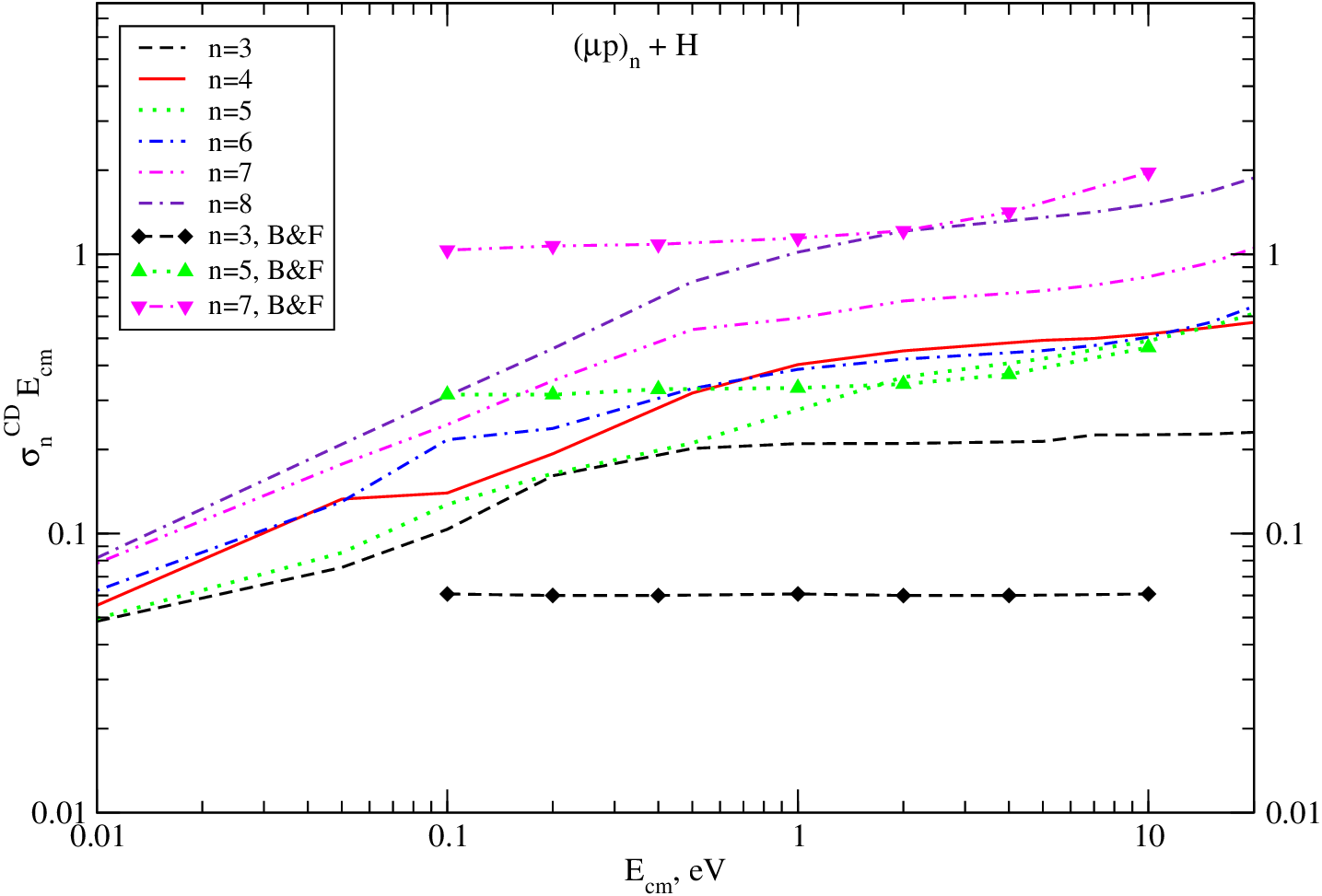}}
     \caption{The cross sections of Coulomb deexcitation
     $\sigma^{\rm CD}_n$ multiplied by $E_{\rm cm}$ for $(\mu p)_n + H$
 collisions calculated in the quantum-mechanical close-coupling method 
 in comparison with the results of Bracci and Fiorentini~[11] (B\&F).}
     \end{figure}

As for the $n$ dependence of the total Coulomb deexcitation
cross sections $\sigma_n^{CD}(E)$ the situation is more
complicated than it is followed from the semiclassical
model~[11] (see Fig.2). There is no dependence
$\sigma_n^{CD}\sim n^{\gamma}\; (\gamma\sim 2)$ as it is
followed from [11]. Our prediction for the $n$ dependence
of the CD cross sections is quite different. In our
opinion, this behaviour is explained by the influence of
the open channels with $\Delta n >1$ which increases for
the higher $n$ states and can`t be taken into account in
the two-state approaches [11-13].

The distribution over the final states $n'$ is completely
different from the semiclassical results [11] as
illustrated in Fig.~4 and Table~3. The present calculations
predict that $\Delta n=1$ transitions dominate in agreement
with the semiclassical, AAA and CTMC results however the
transitions with $\Delta n> 1$ are strongly enhanced as
compared to the results [11-13]. To determine the fraction
of the transitions with $\Delta n>1$ in the total CD cross
section,  the values
\[ f_n=\frac{\sigma^{\rm CD}_n -\sigma_{n,n-1}}{\sigma^{\rm CD}_n}\cdot 100\%.\]
are also presented.

The calculated CD cross sections $\sigma_{n,n-1}(E)$,
$\sigma^{\rm CD}_n(E)$ and $f_n(E)$ for $(\mu p)_n +H$
collisions with $n=3\div 8$ and energies from 0.01 up to
20~eV are presented in Table~3. The observable variations
of the CD cross sections with $E$ can be related to the
opening of the additional channels with $\Delta n>1$. The
violation of the known in literature [11-13] $n$-dependence
of the CD cross sections is illustrated in Figs.~2 and 3
and in more detail in Table~3. For $n=4\div 6$ we observe
the essential variations in the CD cross sections in
contrast to the two-state calculations [11-13]. In Fig.~3
this effect can be clearer because of the different energy
dependence of the CD cross sections for various $n$. In our
opinion the traditional $n$ dependence ($\sigma_n^{\rm CD}
\sim n^{\gamma}$ with $\gamma \ge 2$) followed from the
semiclassical picture can be reached at the relative
energies comparable or much more than $\Delta
\varepsilon_{n,n-1}$.

     \begin{figure}[h]
     \centerline{
     \includegraphics[width=0.7\textwidth,keepaspectratio]{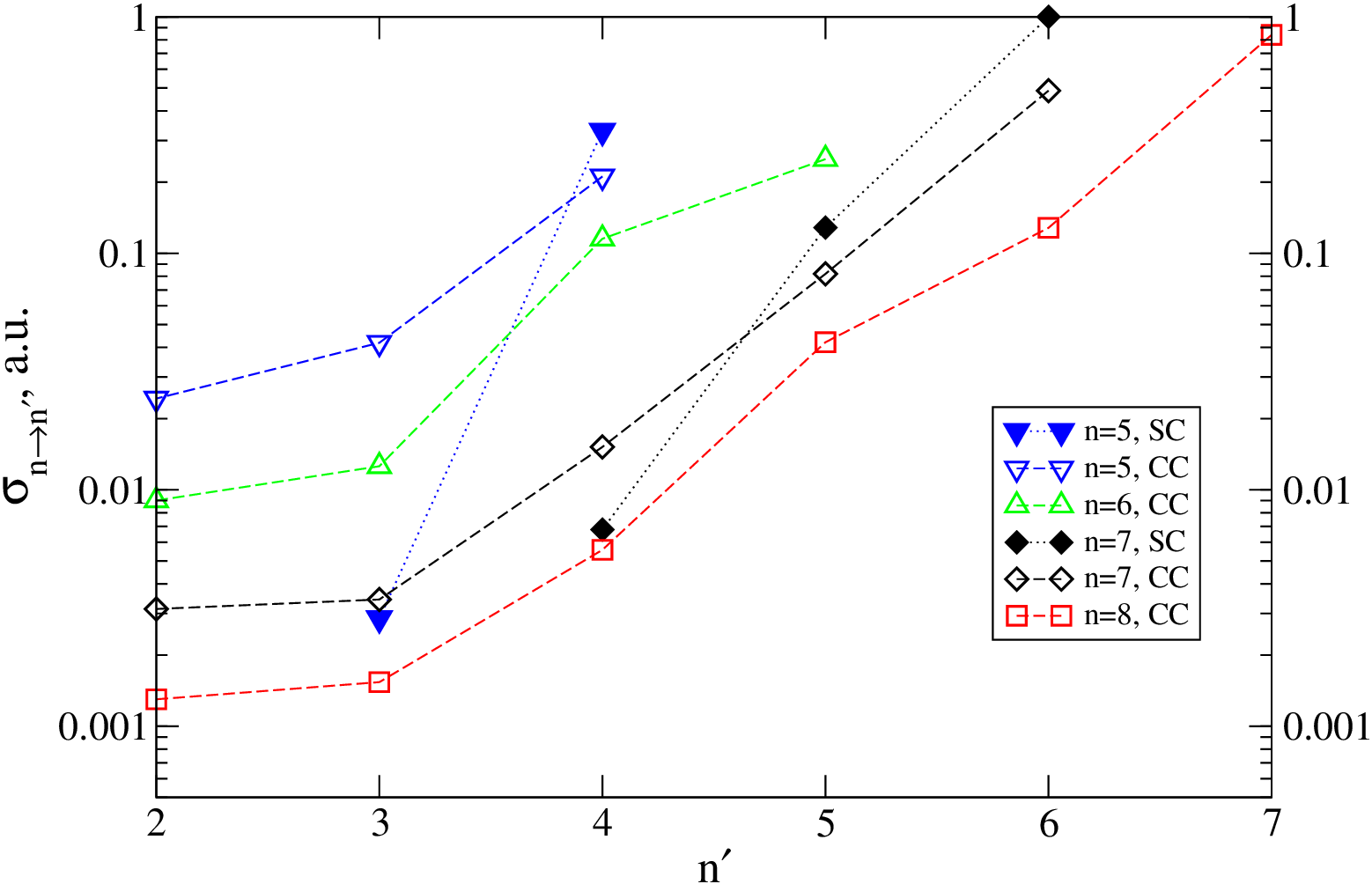}}
     \caption{The Coulomb deexcitation cross sections
     $\sigma_{n,n'}$ in a.u. for $(\mu p)_n + H$
 collisions  at $E=1$~eV calculated in the quantum-mechanical close-coupling method
 in comparison with results of Bracci and Fiorentini~[11].}
     \end{figure}

     \begin{figure}[h!]
     \centerline{
     \epsfig{file=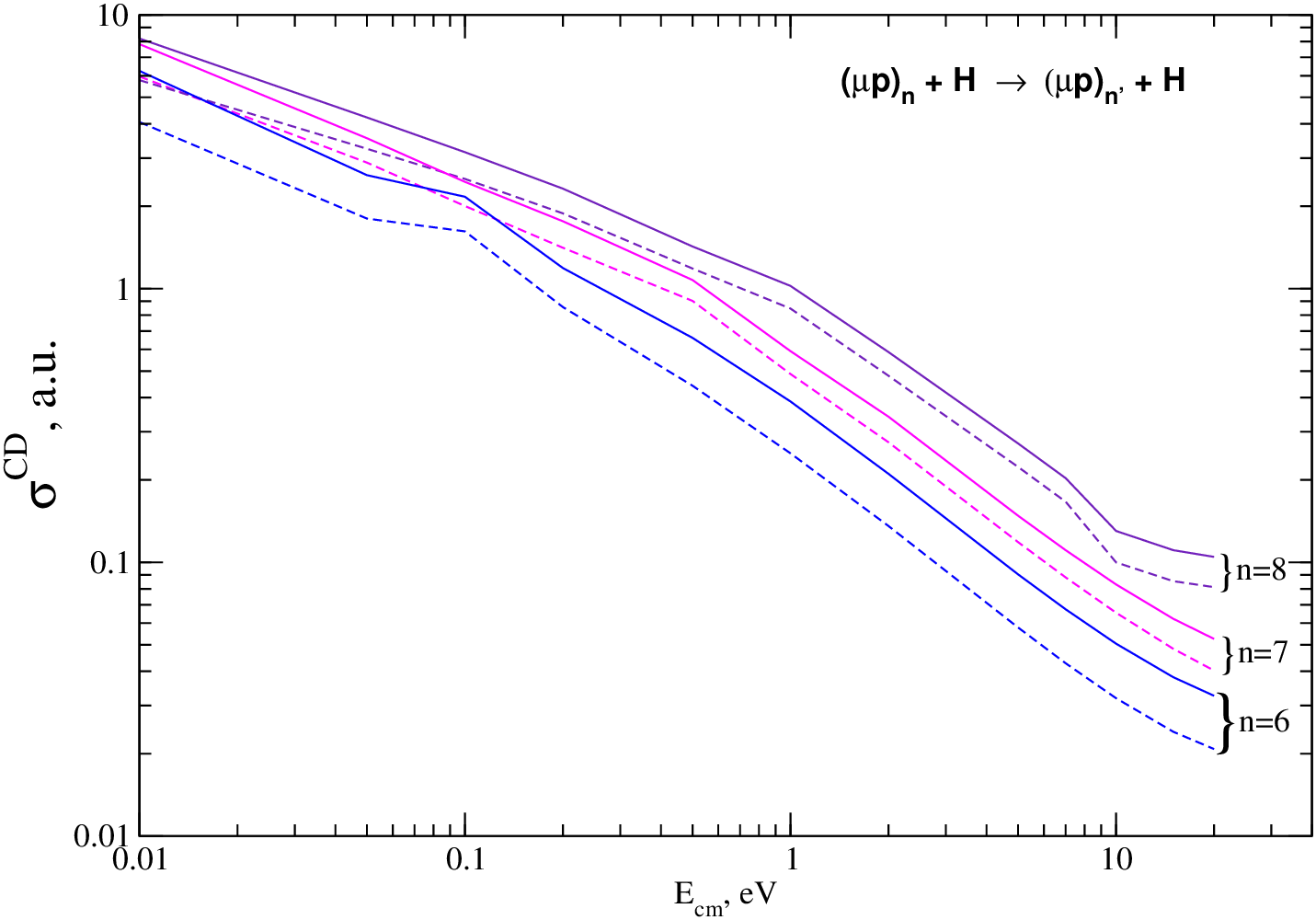,width=0.7\textwidth}}
     \caption{The cross sections of Coulomb deexcitation $\sigma_{n,n-1}$
     (dashed)
  and $\sigma^{\rm CD}_{n}$ (solid) in a.u. for $(\mu p)_n + H$
 collisions calculated in the quantum-mechanical close-coupling method.}
     \end{figure}

The transitions with $\Delta n=1$ are most likely but the
$\Delta n> 1$ transitions make up a substantial fraction of
the total CD cross section (16\% - 37\%) for $n\ge 4$ in
contrast to the two-state approaches [11-13]. The
energy dependence of this fraction is completely different
for various $n$. In particular, this behaviour leads to the
unexpected dependences of the cross sections on the
principle quantum number $n$ and collisional energy $E_{\rm
cm}$.

It is possible that the problem of the high energy fraction
of the kinetic energy distribution in the muonic hydrogen
is related to the unproper behaviour of the CD cross
sections taken in the extended standard cascade model
(ESCM)~[17] for $n=3\div 7$ and based on the semiclassical
model~[11]. The dependences of the CD cross sections on $n$
and $E$ used in ESCM are in disagreement with the results
of our more elaborate close-coupling consideration.

     \begin{figure}[h!]
     \centerline{
     \includegraphics[width=0.6\textwidth,keepaspectratio]{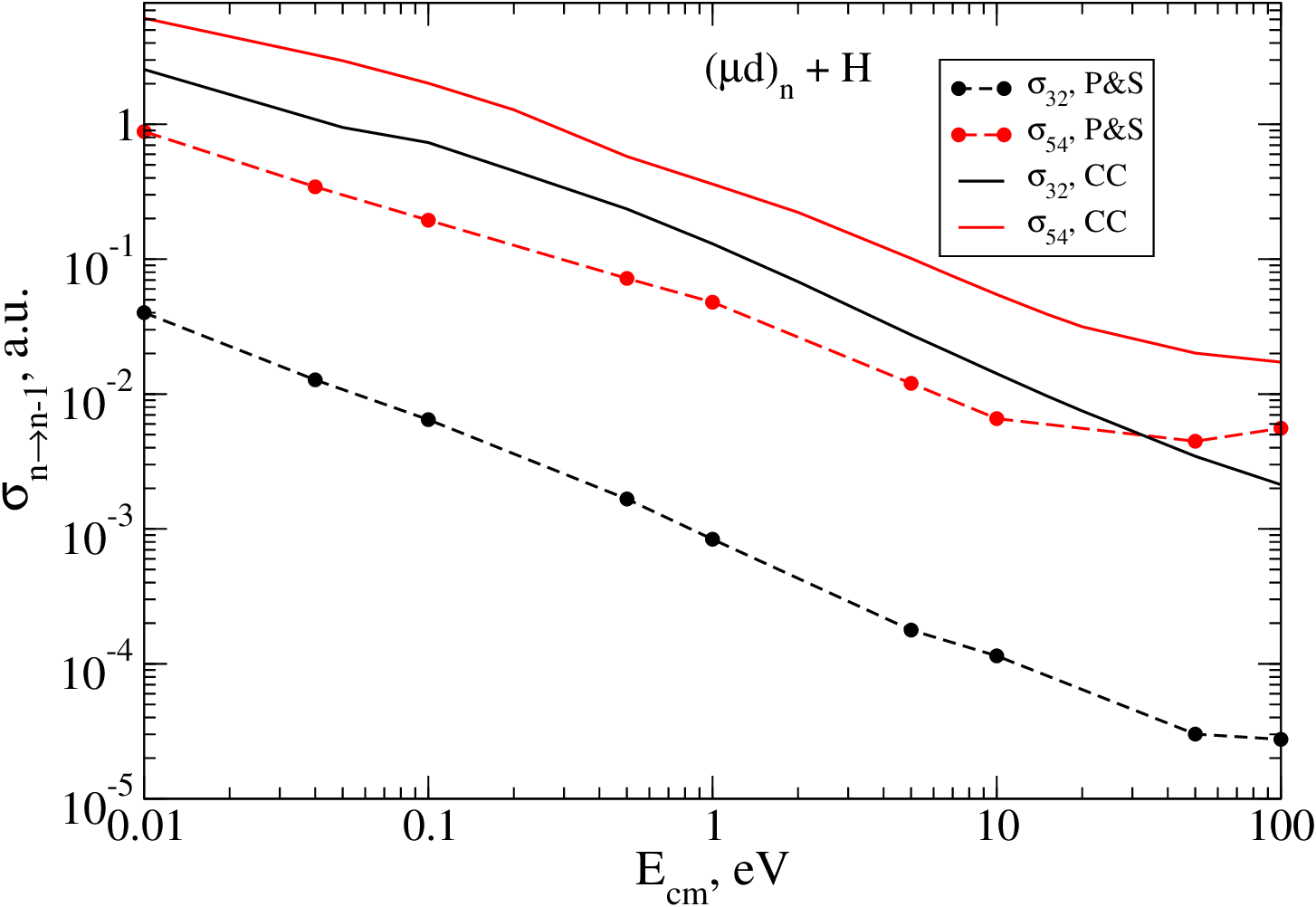}}
     \caption{The cross sections of Coulomb deexcitation
     $\sigma_{n,n-1}$ in a.u. for $(\mu d)_n + H$
 collisions  calculated in the close-coupling approach
 (in two-level approximation) (CC)
 in comparison with results of Ponomarev and Solovyov~[13] (P\&S).}
     \end{figure}

In Fig.~6 we compare the present cross sections of CD
$\sigma_{n,n-1}(E)$ for $(\mu d)_n +H$ collisions with the
results of AAA~[13] for $n=3,5$. As in case of $(\mu p)_n
+H$ [11] the transition $3\to 2$ is strongly suppressed
(almost two order) in comparison with the close-coupling
calculations. At the same time for the transition $5\to 4$
the results of AAA are only suppressed in four times as
compared with our results. The reason of this discrepancy is not clear 
at present.

\section{Conclusion}
The unified treatment of the elastic scattering, Stark
transitions and Coulomb deexcitation is presented in {\em
ab initio} quantum-mechanical approach. The main features
of the Coulomb deexcitation in the collision of the exotic
hydrogen atom in excited states with hydrogen atom have
been investigated in detail using the close-coupling
method. The new results for the $n$ and $E$ dependences of
the CD cross sections are obtained for $n=3\div 8$ and
relative energies up to 20 eV relevant to the kinetics. The
calculated cross sections do not agree with the
long-standing traditional beliefs about $\sigma^{\rm
CD}_n(E)\sim n^{\gamma}/E$  dependence of CD cross sections
which is  based on semiclassical approximation. The new
important results are also obtained for the fraction of the
transitions with $\Delta n>1$. It is shown that the
contribution of these transitions is more essential than it
is assumed earlier and reaches up to $~\sim$37\% for $n=6$.
We suppose that our results provide a reliable theoretical
input for the further kinetics calculations. More detailed
investigations concerning the partial transitions and other
exotic particles will be discussed in  future
publications.\\

{\bf \large Acknowledgments}

We are grateful to Prof. L.Ponomarev for the stimulating
interest and  the fruitful discussions, to the participants
of the Seminar in MUCATEX headed by him for the useful
discussions, to T.Jensen and V.Markushin for the sending
their data on the CD cross sections for comparison with the
present ones and the Russian Foundation for Basic Research
(grant No. 03-02-16616) for financial support.

\end{document}